\documentclass[aps,superscriptaddress,twoside,onecolumn, floatfix,a4paper,longbibliography]{revtex4-2}

\usepackage{graphicx}
\usepackage{amsmath}
\usepackage{dcolumn}   
\usepackage{bm}        
\usepackage{amssymb}   
\usepackage{graphicx,epsfig}
\usepackage{color}
\usepackage[usenames,dvipsnames]{xcolor}
\usepackage[ocgcolorlinks,colorlinks=true,linkcolor=blue,citecolor=red]{hyperref}
\usepackage{amsmath,amssymb, amsthm, amsfonts}
\usepackage{xspace}
\usepackage{xcolor}
\usepackage{verbatim}
\usepackage{mathtools}
\usepackage{lipsum}
\usepackage{natbib}
\definecolor{goodgreen}{rgb}{0.1,0.5,0}

\usepackage[utf8]{inputenc}

\begin{document}
\title{Thermal superconducting quantum interference proximity transistor}
\newcommand{\orcid}[1]{\href{https://orcid.org/#1}{\includegraphics[width=8pt]{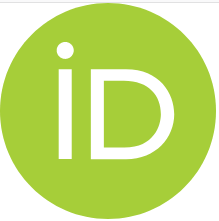}}}

\author{Nadia Ligato}
\affiliation{NEST, Istituto Nanoscienze-CNR and Scuola Normale Superiore, I-56127 Pisa, Italy}
\affiliation{These authors equally contributed to the work.}

\author{Federico Paolucci}
\affiliation{NEST, Istituto Nanoscienze-CNR and Scuola Normale Superiore, I-56127 Pisa, Italy}
\affiliation{These authors equally contributed to the work.}

\author{Elia Strambini}
\email{elia.strambini@sns.it}
\affiliation{NEST, Istituto Nanoscienze-CNR and Scuola Normale Superiore, I-56127 Pisa, Italy}

\author{Francesco Giazotto\orcid{0000-0002-1571-137X}}
\email{francesco.giazotto@sns.it}
\affiliation{NEST, Istituto Nanoscienze-CNR and Scuola Normale Superiore, I-56127 Pisa, Italy}

\maketitle

\textbf{Superconductors are excellent thermal insulators at low temperature owing to the presence of the energy gap in their density of states (DOS) \cite{Giazotto2006}. In this context, the superconducting \textit{proximity effect} \cite{deGennes} allows to tune the local DOS of a metallic or superconducting wire by controlling the phase bias ($\varphi$) imposed across it \cite{Likharev1979}. As a result, the wire thermal conductance can be  tuned over several orders of magnitude by phase manipulation \cite{Giazotto2006}. 
Despite  strong implications in nanoscale heat management \cite{Pop2010}, experimental proofs of phase-driven control of thermal transport in superconducting proximitized nanostructures are still very limited \cite{chandrasekar,eom,petrashov}.
Here, we report the experimental demonstration of efficient heat current control by phase tuning the superconducting proximity effect. This is achieved  by exploiting the magnetic flux-driven manipulation of the DOS of a quasi one-dimensional aluminum nanowire forming a weak-link embedded in a superconducting ring \cite{Giazotto2010,Virtanen2016}. 
Our thermal superconducting quantum interference transistor (T-SQUIPT) shows temperature modulations up to $\sim 16$ mK yielding a temperature-to-flux transfer function as large as $\sim 60$ mK/$\Phi_0$.
Yet, phase-slip transitions occurring in the nanowire Josephson junction induce a hysteretic dependence of its local DOS on the direction of the applied magnetic field. 
Thus, we also prove the operation of the T-SQUIPT as a phase-tunable \textit{thermal memory} \cite{Fornieri2016,Guarcello2018}, where the information is encoded in the temperature of the metallic mesoscopic island.
Besides their relevance in quantum physics, our results are pivotal for the design of innovative devices with application in coherent caloritronics \cite{Fornieri2017,Hwang2020} and energy harvesting \cite{Sothmann2014,Benenti2017}, such as heat valves \cite{Strambini2014}, temperature amplifiers \cite{Paolucci2017} suitable for thermal logic architectures \cite{Paolucci2018}, thermoelectric heat engines \cite{Marchegiani2020,Marchegiani2020_2} and qubits \cite{Iorio2021}.}

\begin{figure*}[th!]
 \centering 
\includegraphics[width=0.80\linewidth]{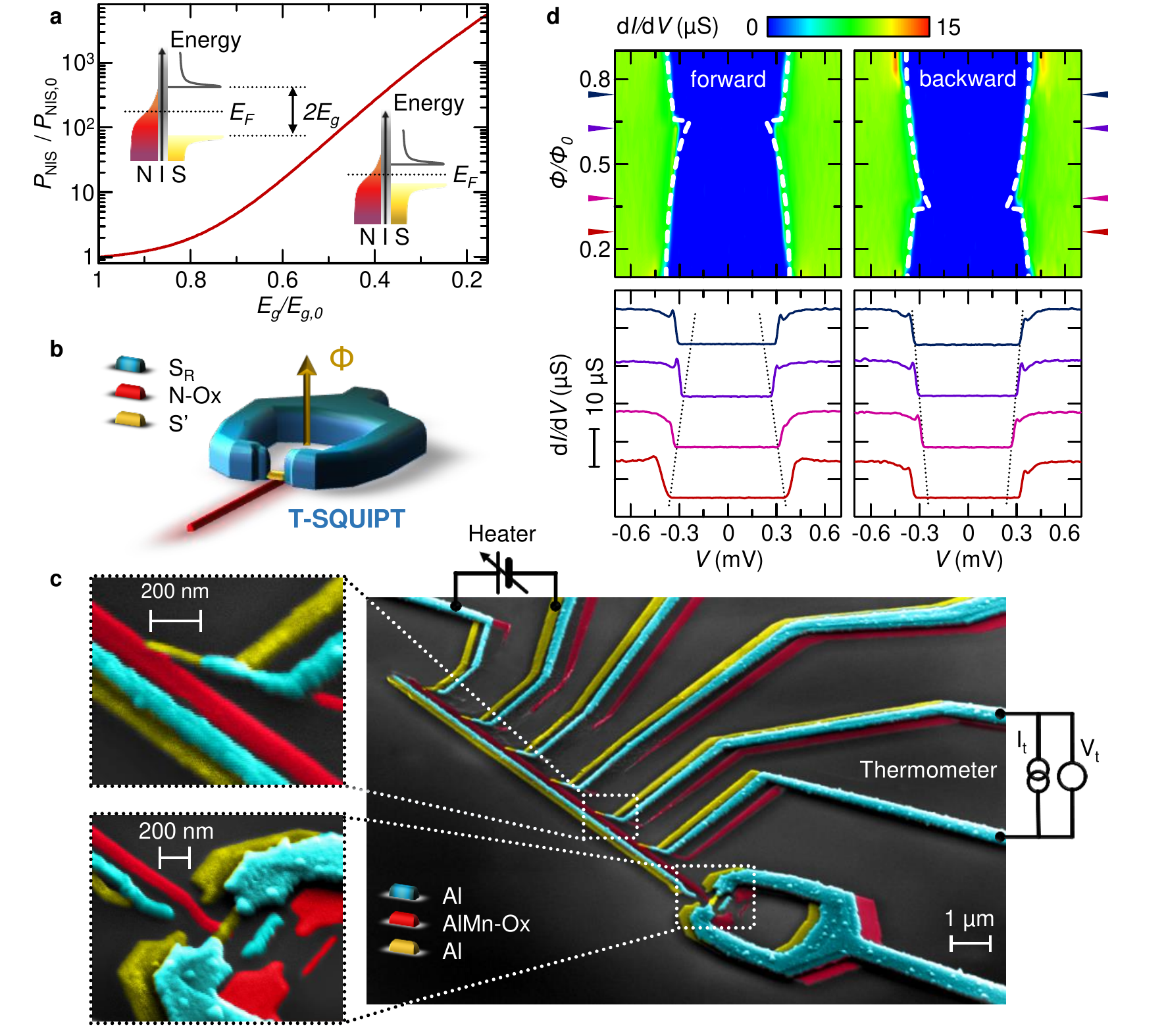}
\caption{\textbf{Thermal superconducting quantum interference proximity transistor operation principle and implementation.}  \textbf{a} Normalized thermal current through a $NIS$ tunnel junction (i.e., {$P_{NIS}$/$P_{NIS,0}$}) versus normalized energy gap of $S$ (i.e., $E_{g}$/$E_{g,0})$. 
$E_{g,0}$ is the maximum value of the energy gap, while $P_{NIS,0}$ is the thermal current flowing for $E_{g,0}$ at constant temperature gradient across the junction. 
The trace is obtained by solving Eq. \ref{NISpower} as a function of $E_g$ considering $T_N=0.1T_{C}$ (where $T_{C}=1.764/E_{g,0}$ is the superconducting critical temperature), $T_S=0.01T_{C}$, and $\Gamma=10^{-4}E_{g,0}$ (see Methods for details). 
A decrease of $E_g$ of about $85\%$ causes an enhancement of $P_{SIN}$ of almost 4 orders of magnitude. Insets: Schemes of the energy band diagram of a temperature biased $NIS$ junction. $E_F$ denotes the Fermi energy.
\textbf{b} Sketch of the thermal superconducting quantum interference transistor (T-SQUIPT). 
A proximitized superconducting nanowire ($S'$, yellow) is phase-biased via a magnetic flux ($\Phi$) piercing a superconducting loop ($S_{R}$, blue). A normal metal electrode ($N$, red) is tunnel coupled to $S'$ through an oxide layer (Ox) thus forming a $NIS'$ junction. The T-SQUIPT is operated by setting different electronic temperatures in $S'$ and $N$.
\textbf{c} False-color scanning electron micrograph of a typical T-SQUIPT. An Al nanowire ($S'$, yellow) is inserted in an Al ring ($S_{R}$, blue), while an Al$_{0.98}$Mn$_{0.02}$ normal metal electrode ($N$, red) is tunnel-coupled through a thin aluminum oxide layer (Ox) to the middle of the nanowire. A set of superconducting aluminum tunnel probes (yellow) are coupled to $N$, and serve as local heaters and thermometers. Insets: blow-up of a thermometer (top) and the $NIS'$ region (bottom).
\textbf{d} (top) Differential conductance of a typical T-SQUIPT measured at $T=25$ mK for a forward (left) and backward (right) sweep of the magnetic flux. The white dashed lines are estimates of $E_{g'}(\Phi)$ obtained with the model presented in the Supplementary Information. (bottom) $\text{d}V/\text{d}I$ versus $V$ measured for the selected value of $\Phi$ indicated by the arrows in the top panels. The differential conductance is hysteretic with respect to
the direction of $\Phi$, due to the multi-valued current-to-phase relation of the long  $S_RS'S_R$ junction.
} 
\label{Fig1}
\end{figure*}

Thermal transport in superconductors stems only from quasiparticle excitations, since Cooper pairs are unable to carry heat \cite{Giazotto2006}. At low temperature, the presence of quasiparticles is exponentially damped  \cite{Tinkham} thanks to the amplitude of the energy gap ($E_g\gg k_B T$, where $k_B$ is the Boltzmann constant and T the temperature) present in the DOS so that superconductors turn out to be excellent thermal insulators \cite{Giazotto2006}. 
Similarly, the heat current ($P_{NIS}$) flowing from a normal metal ($N$, at temperature $T_N$) to a tunnel-coupled superconductor ($S$, at temperature $T_S$) in a temperature-biased $NIS$ junction (where $I$ indicates an insulating tunnel barrier and $T_N>T_S$) is heavily suppressed when the thermal energy is much lower than $E_g$ \cite{Giazotto2006}. For a given temperature gradient, $P_{NIS}$ can rise by several orders of magnitude by lowering $E_g$, as shown in Fig. \ref{Fig1}a (see Methods for details).
Yet, in a proximitized superconductor ($S'$) the superconducting energy gap ($E_{g'}$) can be finely controlled by tuning the macroscopic quantum phase across it ($\varphi$) \cite{deGennes, Likharev1979,Virtanen2016} so that thermal transport through a $NIS'$ tunnel junction is, in principle, $\varphi$-tunable \cite{Fornieri2017,Hwang2020}. The integration of $S’$ into a superconducting loop (see Fig. \ref{Fig1}b) allows to control $\varphi$ \cite{Giazotto2010,Virtanen2016} and thereby to modulate its DOS with an external magnetic flux piercing the ring area ($\Phi$), as already experimentally demonstrated for the electronic transport \cite{Ligato2017,Ronzani2017,Ligato2020}. 
Accordingly, $\Phi$ is expected to master as well the thermal properties of $S'$ \cite{Giazotto2006,Strambini2014}. Indeed, the possibility to $\Phi$-tune the thermal conductance of a proximitized normal metal was demonstrated in Andreev interferometers showing a limited temperature modulation ($\delta T<0.5$ mK) \cite{chandrasekar,eom,petrashov}.
A larger control of the heat transport properties in proximitized systems would enable the development of the thermal counterparts of widespread electronic devices \cite{Fornieri2017,Hwang2020}, such as transistors \cite{Paolucci2017}, logic gates \cite{Paolucci2018} and memories \cite{Fornieri2016,Guarcello2018}.

Here, by exploiting a temperature-biased $NIS'$ junction embedded in a superconducting ring \cite{Giazotto2010,Virtanen2016} we realize a thermal superconducting quantum interference proximity transistor (T-SQUIPT) \cite{Strambini2014} based on the phase-driven manipulation of the heat current ($P_{NIS'}$) via proximity effect.
The implementation of our T-SQUIPT is shown in the false-color scanning electron micrograph presented in Fig. \ref{Fig1}c. The device was realized by standard nano-fabrication techniques employing conventional metals (see Methods for fabrication details). The $S'$ element consists of a 380-nm-long quasi one-dimensional aluminum wire (yellow, $w=90$ nm and $t=20$ nm), while the  superconducting ring ($S_R$) is made of a 70-nm-thick Al layer (blue). The $N$ reservoir (red, 15-nm-thick Al$_{0.98}$Mn$_{0.02}$) is
coupled
to $S'$ via an AlMn-Ox tunnel junction (indicated with Ox in Fig. \ref{Fig1}c) with normal-state resistance $R_T\simeq$ 65 k$\Omega$. To perform thermal transport experiments, a series of superconducting aluminum heaters/thermometers (yellow, 25-nm-thick) \cite{Giazotto2006} are tunnel coupled to $N$ (see Methods for details).



\begin{figure*}[ht!]
	\centering 
	\includegraphics[width=0.8\linewidth]{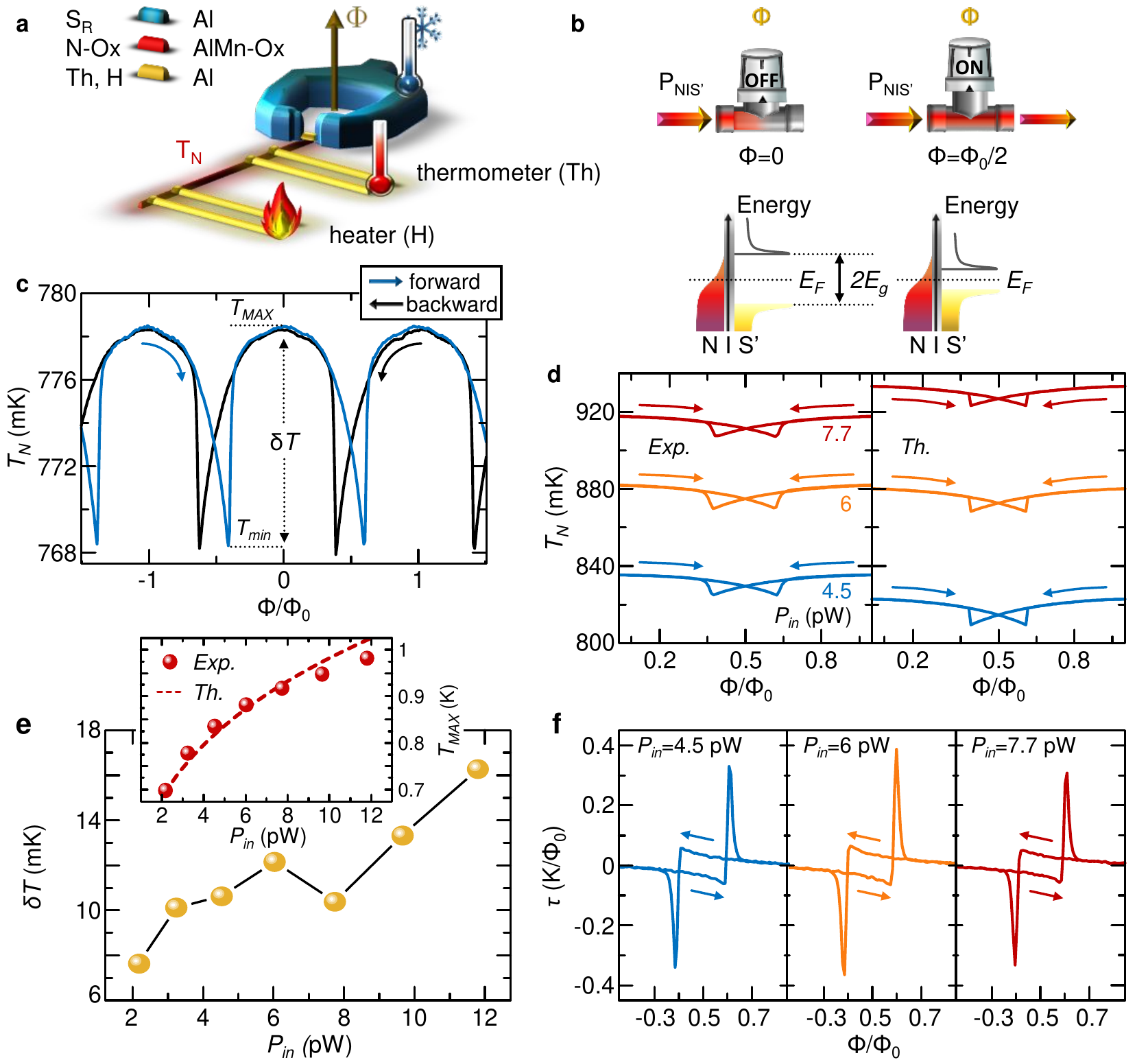}
	\caption{\textbf{Low temperature behavior of the T-SQUIPT.}
	\textbf{a} Sketch of the thermal setup of the T-SQUIPT. The electronic  temperature ($T_N$) in the N electrode (red, Al$_{0.98}$Mn$_{0.02}$) is raised by injecting power through a pair of superconducting heaters, and is measured via a pair of superconducting thermometers. All these $S$ electrodes (yellow, Al) are tunnel-coupled to $N$.
	\textbf{b} Schematic representation of the magnetic flux control of the heat current ($P_{NIS'}$) flowing through the T-SQUIPT. The flux tunes the energy gap of $S'$ ($E_{g'}$): In particular, $E_{g'}(\Phi)$ is maximum for $\Phi=0$ [$P_{NIS'}(0)$ is minimum] and is strongly suppressed for $\Phi=\Phi_0/2$ ($P_{NIS'}$ is enhanced).
	\textbf{c} Electronic temperature [$T_N(\Phi)$] of $N$ acquired for $P_{in}=3.2$ pW. The blue (forward) and black (backward) traces show different magnetic flux sweep directions. The maximum ($T_{MAX}$) and minimum ($T_{min}$) values of $T_N$ together with the temperature swing ($\delta T$) are indicated. 
	\textbf{d} Experimental (left) and theoretical (right) electronic temperature of $N$ vs magnetic flux [$T_N(\Phi)$] for different values of $P_{in}$.
	\textbf{e} Dependence of the electronic temperature swing ($\delta T$) on the injected power ($P_{in}$). $\delta T$ raises with $P_{in}$ within the explored range of heating power. Inset: Maximum electronic temperature ($T_{MAX}$) versus $P_{in}$ obtained from experiments (dots) and theory (dashed line). $T_{MAX}$ monotonically increases with $P_{in}$.
	\textbf{f} Electronic temperature-to-flux transfer coefficient ($\tau$) vs $\Phi$ calculated from the experimental traces shown in panel \textbf{d}. All curves refer to a bath temperature $T_b=25$ mK.}
	\label{Fig2}
\end{figure*}

As demonstrated by the tunneling spectroscopy measurements shown in Fig. \ref{Fig1}d, the T-SQUIPT differential conductance [and therefore $E_{g'}(\Phi)$] is modulated by the external magnetic flux with the expected $\Phi_0$-periodicity (where $\Phi_0\simeq2.067 \times 10^{-15}$ Wb is the flux-quantum), together with a hysteresis associated to the phase-slip transition occurring in \textit{long} $S_RS'S_R$ junctions \cite{Little1967,Likharev1979,Virtanen2016}. Yet, the absence of the conductance peaks around zero bias due to Josephson coupling testifies that the AlMn probe of the T-SQUIPT lies in the normal-state \cite{Ronzani2017}. The detailed charge transport properties of the T-SQUIPT are reported elsewhere \cite{Ligato2020}.

We exploit the same structure to prove the phase-coherent control of heat current $P_{NIS'}(\Phi)$.
Thermal measurements are performed by recording the electronic temperature in $N$ [$T_N(\Phi)$] while heated by a constant power $P_{in}$ at fixed bath temperature ($T_b$), as depicted in Fig. \ref{Fig2}a (see Methods for details). The resulting temperature gradient set across the $NIS'$ junction gives rise to a finite thermal current $P_{NIS'}(\Phi)$, which is the only $\Phi$-dependent heat exchange channel for $N$ (see Methods for details). As a consequence, the variation of $T_N$ with the magnetic flux arises only from the modulation of $E_{g'}(\Phi)$ via the proximity effect, as sketched in Fig. \ref{Fig2}b. 

Figure \ref{Fig2}c shows $T_N$ as a function of $\Phi$ measured for $P_{in}=3.2$ pW at $25$ mK. Stemming from  $\Phi$-modulation of the nanowire DOS (and thus $P_{NIS'}$), $T_N$ is $\Phi_0$-periodic and hysteretic with the sweep direction of $\Phi$, similarly to the electronic counterpart \cite{Ligato2020}. The strong suppression of $P_{NIS'}$ occurring for $\Phi=0$ [where 
$E_{g'}(0)$ is maximal] yields the maximum steady-state electronic temperature established in $N$ ($T_{MAX}$), while  thermal current enhancement at higher magnetic fluxes causes the decrease of $T_N$ down to its minimum ($T_{min}$). Notably, $T_{min}$ is not recorded exactly at $\Phi=0.5\Phi_0$, and  is hysteretic in the magnetic flux as a consequence of the phase-slip transitions occurring in $S'$ \cite{Little1967,Likharev1979,Virtanen2016,Ligato2020}. The width of the hysteresis loop is exclusively due to the phase dependence of the nanowire free energy, since its temperature is poorly affected by $T_N$ due to the good thermal contact with the substrate and the $S$ ring (see Supplementary Information for details). Moreover, the probability of unwanted phase-slip events arising from quantum tunneling \cite{Mooij2005} or thermal activation \cite{Arutynov2008} is almost zero for our device \cite{Ligato2020}.

Let us now investigate the impact of $P_{in}$ on the performance of the T-SQUIPT. Figure \ref{Fig2}d shows the $T_N(\Phi)$ characteristics measured and calculated at $T_b=25$ mK for three selected values of $P_{in}$ (see Methods for details). Although idealized, our model provides a reasonable agreement with the experimental data, thereby grasping the relevant physical picture at the origin of the effect in our device. At a given $\Phi$, the value of $T_N$ monotonically rises by increasing the input power, as shown in the inset of Fig. \ref{Fig2}e. Similarly, the temperature modulation ($\delta T$) grows with the input power. In particular, the T-SQUIPT shows a maximum $\delta T\sim 16$ mK for $P_{in}=12$ pW, which corresponds to a relative variation of $T_N$ of about $1.7\%$. 

The flux-to-temperature transfer coefficient ($\tau=\partial T_N/\partial \Phi$) is a key parameter to evaluate the response of the T-SQUIPT. 
Figure \ref{Fig2}f shows $\tau(\Phi)$ for different values of $P_{in}$. 
Due to the highly non-linear dependence of $P_{NIS'}$ on the superconducting energy gap (see Fig. \ref{Fig1}a), $\tau$ is vanishing for low phase-biases, and it increases up to about 60 mK/$\Phi_0$ for flux values approaching $0.5\Phi_0$. This value is on par with the first realization of a Josephson heat interferometer \cite{Giazotto2012}. 
We also note that the T-SQUIPT shows similar values of $\tau$ for the input powers selected in Fig. \ref{Fig2}f, since $\delta T$ is almost constant in the range 3.5 pW $\leq\;P_{in}\;\leq$ 8 pW (see Fig. \ref{Fig2}e).
Furthermore, the transfer coefficient shows also a peak corresponding to the abrupt temperature change occurring at the phase-slip transition. This feature of $\tau$ cannot find practical applications, since it depends on experimental technicalities, such as the spacing and the speed of the magnetic flux sweep. 


\begin{figure}[t!]
	\includegraphics{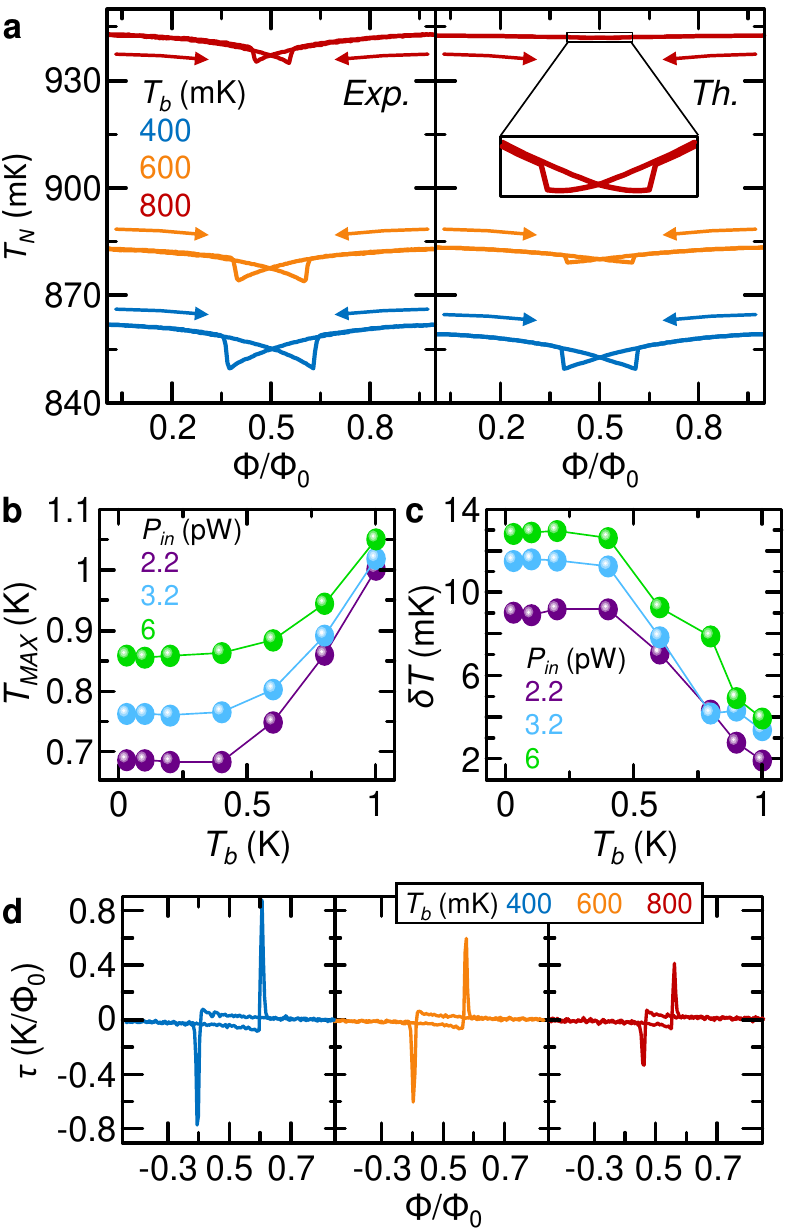}
	\caption{\textbf{Bath temperature evolution of the T-SQUIPT behavior.} 
	\textbf{a} Experimental (left) and theoretical (right) electronic temperature modulation [$T_N(\Phi)$] of the T-SQUIPT at $P_{in}=6$ pW for different values of $T_b$. The theoretical curve at $T_b=600$ mK (red) is blown up to highlight the presence of hysteresis. The model well describes the data to bath temperatures up to 600 mK. At higher values of bath temperature, the model underestimates the flux-dependent temperature modulation.
	\textbf{b} Maximum electronic temperature in $N$ ($T_{MAX}$) versus $T_b$ for selected values of input power. For each $P_{in}$, $T_{MAX}$ monotonically rises with $T_b$.
	\textbf{c} Bath temperature dependence of $\delta T$ for selected values of $P_{in}$. The swing of $T_N$ is rather suppressed for $T_b \gtrsim 400$ mK. 
    \textbf{d} Flux-to-temperature transfer coefficient versus $\Phi$ at $P_{in}=6$ pW for the same values of $T_b$ as in panel \textbf{a}.}
	\label{Fig3}
\end{figure}

We now discuss the impact of bath temperature on the performance of the T-SQUIPT. Figure \ref{Fig3}a displays the experimental (left) and theoretical (right) $T_N(\Phi)$ characteristics of the heat nanovalve for $P_{in}=6$ pW and different values of $T_b$. At a given input power, $T_{MAX}$ starts to rise for $T_b\geq400$ mK, as shown in \ref{Fig3}b. Moreover, the temperature difference between the forward and backward traces is reduced by increasing $T_b$, since the electron-phonon interaction in $N$ becomes larger at higher bath temperatures \cite{Giazotto2006,Wellstood1994}. At the same time, $\Phi$-tuning of $P_{NIS'}$ is less effective at high $T_b$, because the energy gap of $S'$ is lowered while approaching its critical temperature \cite{deGennes} (see Methods for details).
Therefore, by rising $T_b$ the temperature swing $\delta T$ for given $P_{in}$ is damped to $\sim 20\%$ of its low temperature value (see Fig. \ref{Fig3}c). Notably, the transfer  coefficient $\tau$ is only slightly affected by the decrease of $\delta T$. As a matter of fact, $\tau(\Phi=0.5\Phi_0)$ remains almost constant by increasing the bath temperature, as shown in Fig. \ref{Fig3}d. By contrast, the hysteresis in the $T_N$ versus $\Phi$ traces is reduced by raising $T_b$, since the $S'$ coherence length [$\xi_{S'}(T)\propto E_g(T)^{-1/2}$] becomes larger at higher temperatures \cite{Tinkham} (see Methods for details) thereby causing the nanowire transition towards the short-junction limit \cite{Likharev1979,Virtanen2016}.

\begin{figure*}[ht!]
	\centering 
	\includegraphics[width=0.9\linewidth]{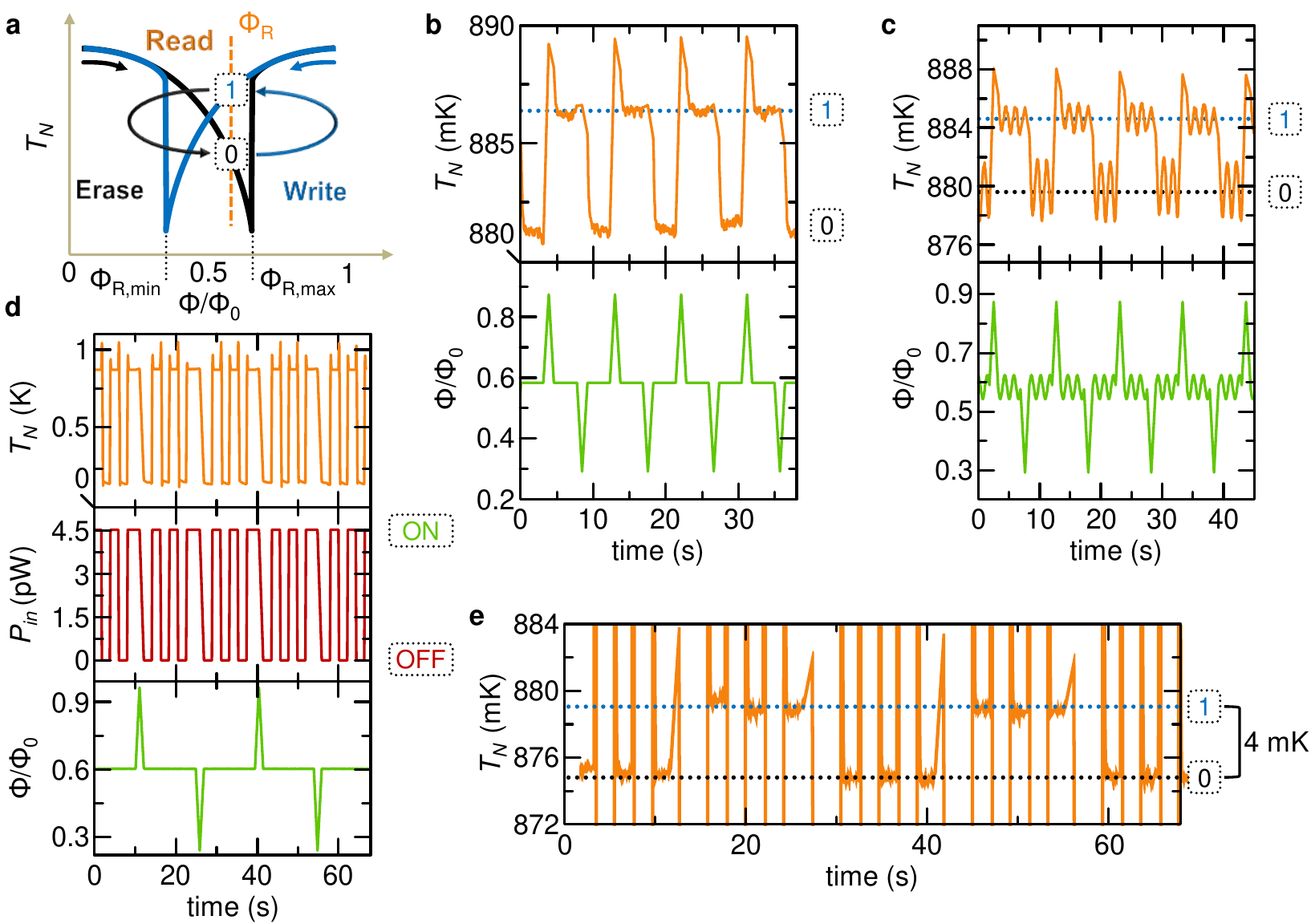}
	\caption{\textbf{T-SQUIPT operated as thermal memory cell.} 
	\textbf{a} Schematic representation of the T-SQUIPT operation as thermal memory cell. The logic states are encoded in low $T_{[0]}$ (black) and high $T_{[1]}$ (blue) temperature branches present at the read-out flux $\Phi_R \in (0.5\Phi_0,\Phi_{R,max})$. The write operation is performed by applying $\Phi_W > \Phi_{R,max}$, while the erase procedure consists on flux biasing the T-SQUIPT at $\Phi_E < \Phi_{R,min}$.
	\textbf{b} Dependence of the read-out electronic  temperature $T_N$ (top panel) on the magnetic flux (bottom panel) measured for $P_{in}=4.5$ pW. The read-out flux is $\Phi_R=0.58\Phi_0$, the write pulse reaches $\Phi_W=0.87\Phi_0$ and the erase flux is $\Phi_E=0.29\Phi_0$.
	\textbf{c} Dependence of the read-out electronic temperature $T_N$  (top panel) on the magnetic flux (bottom panel) measured for $P_{in}=4.5$ pW. The read-out flux is the sum of the constant component $\Phi_R=0.58\Phi_0$ and a sinusoidal oscillation $\Phi_{AC}= \pm 0.04\Phi_0$. The write and erase fluxes are $\Phi_W=0.87\Phi_0$ and $\Phi_E=0.29\Phi_0$, respectively.
	\textbf{d} Dependence of the read-out temperature (top panel) on the magnetic flux (bottom panel). The T-SQUIPT is powered (at $P_{in}=4.5$ pW, central panel) only during the read-out operation (at $\Phi_R=0.58\Phi_0$). The write and erase fluxes are $\Phi_W=0.96\Phi_0$ and $\Phi_E=0.21\Phi_0$, respectively. Notice that at $P_{in}=0$, we recorded $T_N\simeq150$ mK instead of 25 mK, since the SINIS thermometer \cite{Giazotto2006} was calibrated to be sensitive in the temperature range where the T-SQUIPT shows sizable $\Phi$-dependent modulations (i.e., for $T_N>500$ mK). \textbf{e} Zoom of the read-out electronic temperature $T_N$ around state [0] and state [1] measured for $P_{in}=4.5$ pW. The difference between the two memory states is $\sim 4$ mK. All data were recorded at $T_b=25$ mK.}
	\label{Fig4}
\end{figure*}

Our T-SQUIPT may find application as a phase-tunable memory element suitable for {\it thermal computing} \cite{Paolucci2018}. Indeed, in full analogy with superconducting electronic memories \cite{Ligato2020,Baek2014,Golod2015,Gingrich2016,Ligato2020}, the hysteretic $T_N(\Phi)$ characteristics can be exploited to store logic information, as schematically shown in Fig. \ref{Fig4}a. 
In the T-SQUIPT, the two distinct temperature states (i.e., \textit{low} [0] and \textit{high} [1]) at the read out flux ($\Phi_R$) are characterized by a different {\it topological} index, and can be discriminated by the parity of the winding number of $\varphi$ along the Al wire (\textit{odd} or \textit{even}) \cite{Little1967,Ligato2020}.
These states are only accessible by sweeping the magnetic flux in opposite directions, as demonstrated in the electrical counterpart \cite{Ligato2020}. 
Indeed, writing  (erasing) the T-SQUIPT is performed by temporarily rising (lowering) the total flux above (below) the hysteresis loop. Unwanted stochastic transitions (due, for instance, to quantum tunneling or thermal activation) are prevented in the T-SQUIPT by the large phase-slip energy barrier \cite{Little1967} present in the wire thereby making the phase-slip rate vanishing up to about the superconducting critical temperature \cite{Ligato2020}. This enables the use of the T-SQUIPT as a {\it persistent thermal}  memory cell. 


The primary thermal memory cell operation is based on a continuous read-mode, which requires a permanent power injection $P_{in}$ in the N region. In Fig. \ref{Fig4}b, the device is powered with $4.5$ pW and the read-out flux is set above the crossing point of the two temperature branches ($\Phi_R=0.58\Phi_0$). Logic state [1] is written by applying a magnetic flux pulse $\Phi_W=0.87\Phi_0$, and corresponds to $T_N\simeq886$ mK. The erasing operation is implemented by a counter pulse $\Phi_E=0.29\Phi_0$, and results in $T_N\simeq881$ mK (i.e., logic state [0]). Notably, the T-SQUIPT writing/erasing operations were performed several times with high reproducibility.

Stability with respect to magnetic flux fluctuations is one of the key parameters of a memory cell. Here, the stability is investigated by superimposing a sinusoidal oscillation ($\Phi_{AC}=0.04\Phi_0$) onto the read-out flux, as shown in Fig. \ref{Fig4}c for $P_{in}=4.5$ pW and $\Phi_R=0.58\Phi_0$. Despite the flux oscillations amplitude is $\sim 80\%$ of the difference between $\Phi_R$ and the minimum writing flux ($\Phi_{R,max}=0.63\Phi_0$), the stored state is fully preserved in time. As a consequence, we can conclude that the T-SQUIPT is a thermal memory element robust against magnetic fluctuations.

Finally, a \textit{non-volatile} memory retains the stored state when power is temporarily switched off. This property is crucial for energy saving and to design long-term persistent storage units. 
The operation of the T-SQUIPT as a thermal memory requires a source to generate the read-out flux $\Phi_R$ and a power generator for the thermal bias. Despite we provide $\Phi_R$ with an external superconducting coil, the read-out flux could be generated by a permanent ferromagnet without the need of an external source. Therefore, the non-volatility of the T-SQUIPT is simply investigated by turning on the heat input power only during the read-out operation, as shown in Fig. \ref{Fig4}d. 
On the one hand, the steady-state electronic temperature in $N$ is expected to be $T_N=T_b=25$ mK for $P_{in}=0$. Instead, we recorded $T_N\simeq150$ mK. We note that this discrepancy arises only from the thermometer calibration. Indeed, the SINIS thermometer \cite{Giazotto2006} is calibrated to properly operate at $T_N>500$ mK (where the T-SQUIPT shows the interesting flux-dependent temperature modulations) and, therefore, it provides an almost temperature-independent signal for $T_N<150$ mK.
On the other hand, $T_N$ can acquire two distinct values (corresponding to logic state [0] or [1]) when the T-SQUIPT is powered (see top panel of Fig. \ref{Fig4}d and Fig. \ref{Fig4}e). Furthermore, the stored data is not deteriorated by de-powering the memory cell. Indeed, the temperature values corresponding to state [0] ($T_{[0]}\simeq875$ mK) and [1] ($T_{[1]}\simeq879$ mK) remain constant in time after many powering/de-powering cycles (see Fig. \ref{Fig4}e). Therefore, the T-SQUIPT can be, in principle,  employed as a non-volatile memory cell.

In conclusion, we demonstrated phase-tuning of the thermal properties of a nanosized superconductor via the control of its density of states due to the proximity effect. 
In particular, our T-SQUIPT allows a fine magnetic-flux-control of the heat exchange between a proximitized superconducting nanowire and a tunnel-coupled normal metal electrode. 
In good agreement with our theoretical model, the T-SQUIPT shows electronic temperature modulations up to $\sim 16$ mK, and flux-to-temperature transfer functions reaching $\sim60$ mK/$\Phi_0$ around $0.5\Phi_0$. The nanovalve operates up to 1 K showing a reduction of temperature swing of $\sim 80 \%$. 
In addition, the $T_N(\Phi)$ characteristics show also hysteresis, owing to the strong activation energy for the nucleation of a phase-slip in the proximitized nanowire.
This prevents unwanted phase-slips to occur making the T-SQUIPT a prototypical thermal memory cell, where the logic state is encoded in the electronic temperature $T_N$. Our memory element showed robustness against magnetic flux fluctuations and non-volatility, as its electrical counterpart \cite{Ligato2020}. 

From a fundamental point of view, our methods in principle also enable the investigation of chargeless modes in solid state systems that are not accessible by conventional electronic transport, such as Majorana bound states \cite{Ben2015}, topological states \cite{Bours2019,Scharf2020} and parafermions \cite{Banerjee2018}. From the application side, thermal control of heat currents via proximity effect might be improved by exploiting different materials configurations. For instance, the use of \textit {short} Josephson weak-links could enhance the on/off ratio of the thermal valve \cite{Strambini2014}, since the modulation of $\varphi$ would be more efficient \cite{Virtanen2016,Ronzani2014}. Yet, the T-SQUIPT operation might be pushed above liquid helium temperature by exploiting interferometers made of larger gap superconductors, such as vanadium \cite{Ligato2017} or niobium \cite{Jabdaraghi2016}. 
Despite the relevance in nanoscale heat management \cite{Giazotto2006,Pop2010} and thermodynamics \cite{Benenti2017}, the T-SQUIPT could find application in coherent caloritronics \cite{Strambini2014,Fornieri2017,Hwang2020} and energy harvesting \cite{Sothmann2014,Benenti2017}. For instance, it might perhaps be at the core of temperature amplifiers \cite{Paolucci2017}, heat logic architectures \cite{Paolucci2018}, non-volatile data storage units \cite{Fornieri2016,Guarcello2018}, thermoelectric heat engines \cite{Marchegiani2020,Marchegiani2020_2} or qubits \cite{Iorio2021}.


\section*{Methods}
\subsection*{Thermal current through a temperature-biased NIS tunnel junction}
The thermal current flowing form the N to the S electrode of a temperature-biased normal metal/insulator/superconductor ($NIS$) tunnel junction reads \cite{Giazotto2006}
\begin{align} 
P_{NIS}=\frac{1}{e^{2} R_{T}} \int_{-\infty}^{+\infty} \mathrm{d}E  E\;\mathcal{N}_S (E,T_S)  \notag \\ 
\times \left[f_{0}(E,T_{N})-f_{0}(E,T_S)\right],
\label{NISpower}
\end{align}
where $e$ is the electron charge, $R_T$ is the normal-state tunnel resistance, $T_N$ is the temperature of the normal metal, $T_S$ is the temperature of the superconductor, and
$f_{0}(E,T_i)=\left[1+\exp{\left(E/k_{B}T_i)\right)}\right]^{-1}$
is the Fermi-Dirac distribution at temperature $T_i$ (with $i=N,S$).
The normalized density of states of the superconductor takes the form 
$\mathcal{N}_S (E,T_S)=\text{Re}\big[(E+i\Gamma)/{\sqrt{(E+i\Gamma)^2-E_g^2(T_S)}}\big]$, with $\Gamma$ the Dynes parameter accounting for broadening \cite{Dynes1984} and $E_g(T_S)$ the superconducting pairing potential.
The plot in Fig. \ref{Fig1}a is obtained by solving Eq. \ref{NISpower} with $R_T=1$ k$\Omega$, $T_N=0.1T_{C}$ ($T_{C}$ is the superconducting critical temperature corresponding to $E_{g,0}=200\;\mu$eV), $T_S=0.01T_{C}$, and $\Gamma=10^{-4}E_{g,0}$.

\subsection*{T-SQUIPT fabrication}
The T-SQUIPTs were fabricated by a single electron-beam lithography (EBL) step and three-angle shadow-mask metals deposition through a suspended bilayer resist mask. The evaporation and oxidation processes were performed in an ultra-high vacuum (UHV) electron-beam evaporator (base pressure of $10^{-11}$ $\textrm{Torr}$). The use of the same material for the deposition of the nanowire and the ring allows to obtain a transparent interface between them. Furthermore, the room temperature exposure of aluminum films to O$\textsubscript{2}$ produces high quality tunnel junctions. Therefore, the proximitized nanowire, thermometer/heater leads, and the ring of the T-SQUIPTs were all made of Al. At first, 15 nm of Al$\textsubscript{0.98}$Mn$\textsubscript{0.02}$ was deposited at an angle of -18$^\circ$ to realize the N electrode. The volume of the N lead is $\mathcal{V}_N\sim1.7\times10^{-20}$ m$^{3}$.
Subsequently, the sample was exposed to 60 mTorr of $\textrm O_{2}$ for 5 min in order to realize the AlOx thin layer forming the tunnel barriers. Then, a 20-nm-thick Al layer was deposited at a tilt angle of $10^\circ$ to form the superconducting nanowire and tunnel leads. As final step, 70 nm of Al was deposited at $0^\circ$ to realize the superconducting ring (circumference $\sim7.6\;\mu$m, and average width $w_{R,ave}\simeq600$ nm).

\subsection*{Measurement set-up}
The magneto-electric and thermal measurements of the T-SQUIPTs were performed in a $^{3}$He-$^{4}$He dilution refrigerator (Triton 200, Oxford Instruments) equipped with RC-filters (R = $\sim$ 2k$\Omega$) in a temperature range from 25 mK to 1 K. The magnetic field piercing the ring was applied by a superconducting magnet driven by a low-noise current source (Series 2600, Keithley Instruments). 
The preliminary electric characterization was  performed in a two-wire voltage-bias configuration by means of a low-noise DC voltage source (GS200, Yokogawa) and a room-temperature current preamplifier (Model 1211, DL Instruments).
The SINIS thermometers \cite{Giazotto2006} were biased with a current of 150 pA through a battery-powered floating source, while the heaters were piloted upon voltage biasing varied from 900 $\mu$V to 2.1 mV (GS200, Yokogawa), corresponding to about 2.2-11.8 pW injected power range. 
The SINIS thermometers were preliminary calibrated against a RuO$_2$ thermometer anchored to the mixing chamber plate of the fridge while changing the bath temperature.
The voltage drop across the junctions was measured with a standard room-temperature low-noise differential preamplifier (Model 1201, DL Instruments).

\subsection*{T-SQUIPT parameters} \label{sec:Measurements}
The zero-temperature coherence length of the superconducting proximitized wire is $\xi\textsubscript{S',0}$ = $\sqrt{\hbar D/E\textsubscript{g',0}} \simeq 80$ nm, where $\hbar$ is the reduced Planck constant, ${D} ={22.5}$ cm$\textsuperscript{2}$s$\textsuperscript{-1}$ is the diffusion coefficient of the aluminum film, and $E\textsubscript{g',0}\simeq$ 200 $\mu$eV is the Al zero-temperature energy gap. The $S'$  critical temperature is $T_{C,S'}=E\textsubscript{g',0}/1.764k_B\simeq1.31$ K, where $k_B$ is the Boltzmann constant.

At low temperature, the ratio ${L}/ \xi\textsubscript{S',0}\simeq 4.6$, where $L=380$ nm is the nanowire physical length. As a consequence, the T-SQUIPT operates in the long junction limit \cite{Likharev1979}. Non hysteretic behavior of the T-SQUIPT is expected for $\xi\textsubscript{S',short}\gtrsim L/3.5\sim 105$ nm \cite{Likharev1979} providing a temperature upper boundary $T_{single}=T_{C,S'}(1-0.852^2\xi\textsubscript{S',0}l/\xi\textsubscript{S',short}^2)\sim 1.29$ K \cite{Tinkham,Likharev1979}, where $l=3D/v_F\simeq3.3$ nm is the aluminum film mean free path, and $v_F=2.03\times10^6$ m/s is the Fermi velocity of Al.
\
\subsection*{Thermal model of the T-SQUIPT}
The model of the T-SQUIT thermal behavior requires to take into account all the heat exchange mechanisms in the normal metal island. In particular, it is necessary to consider the injected power ($P_{in}$), the losses through the thermometers $P_{th}$, the thermalization with the phonons ($P_{e-ph,N}$) and the flux-dependent transport towards $S'$ [$P_{NIS'} (\Phi)$]. Furthermore, in the T-SQUIPT the electronic temperature in $N$ is non-uniform and the coupling between the AlMn and the substrate phonons is finite (thus $T_{ph,N}\geq T_b$). We can simplify the model by considering a small portion of $N$ included between the thermometers and $S'$ where we can assume a uniform temperature. 
The temperature $T_N$ of this small portion rises thanks to the power injected from the rest of the $N$ island ($P_{in}^*$). $P_{in}^*$ is lower then the total power injected by the heater ($P_{in}$) and, for simplicity, it  will be re-normalized by a damping factor $d_{loss}$ ($P_{in}^*=P_{in}/d_{loss}$). This re-normalisation represents all the thermal losses in the rest of the island ($P_{th}$ and $P_{e-ph}$). Therefore, the thermal balance equation for the restricted island can be simplified to:
\begin{equation}
    \begin{cases}
    P^*_{in}=P^*_{e-ph,N}+P_{NIS'}(\Phi)\\
    P^*_{e-ph,N}=P^*_K,
  \end{cases}
  \label{BalanceTot*}
\end{equation}
where $P^*_K=\alpha A^*_N (T_{ph,N}^4 -  T_b^4)$ (with $\alpha$ the coupling constant and $A^*_N$ the contact area) is the Kapitza coupling between the film and the substrate phonons and the symbol $^*$ refers to the small portion of $N$ that we consider. The simulation of $P_{NIS'} (\Phi)$ requires to know the $E_{g'}(\Phi)$ characteristics (see Eq. \ref{NISpower}). The latter is extracted from the experimental  dependence of $T_N$ on $E_{g'}$. In the above equation, $P^*_{e-ph,N}=\Sigma_N\mathcal{V}^*_N(T_N^6-T_{ph,N}^6)$ is the heat current flowing between electrons and phonons in $N$ \cite{Giazotto2006}, with $\Sigma_N=4.5\times 10^9$ WK$^{-6}$m$^{-3}$ the electron-phonon coupling constant of Al$_{0.98}$Mn$_{0.02}$ and $\mathcal{V}^*_N=0.00375\; \mu$m$^3$ is the volume of the small portion. For the plots of Figs. \ref{Fig2} and \ref{Fig3}, we employed  $A^*_N=0.25\; \mu$m$^2$, $\Gamma=10^{-3}$, $R_T = 65$ k$\Omega$ (extracted from the electrical characterization of the junction), $d_{loss} = 7$ and $\alpha=6.25$ pW$\mu$m$^{-2}$K$^{-1}$. The details of the model and the justification of all the approximations are provided in the Supplementary Information.

\section*{Data availability}
All other data that support the plots within this paper and other findings of this study are available from the corresponding author F.G. upon reasonable request.

\section*{Code availability}
The codes that support the findings of this study are available from the corresponding author E.S. upon reasonable request.

\section*{Acknowledgements}
We acknowledge the European Research Council under Grant Agreement No. 899315-TERASEC, and  the  EU’s  Horizon 2020 research and innovation program under Grant Agreement No. 800923 (SUPERTED) 
and No. 964398 (SUPERGATE)
for partial financial support.

\section*{Author contributions} \label{sec:Author contributions}
E.S. and F.G. conceived the experiment. N.L. fabricated the samples with inputs of F.P.. N.L., F.P. and E.S. performed the measurements. N.L. analysed the experimental data with inputs from F.P., E.S., and F.G.. E.S. and F.G. developed the theoretical model. N.L and F.P. wrote the manuscript with inputs from all authors. All authors discussed the results and their implications equally at all stages.

\section*{Additional information} \label{sec:Additional information}
The authors declare no competing interests.
Supplementary information is available for this paper at https:XXX
Correspondence and requests for materials should be addressed to E.S. or F.G.



\section*{Supplementary Information}

\section*{Thermal model of the T-SQUIPT}
\subsection{Ideal phonon thermalization}

\begin{figure*}[th!]
\centering 
\includegraphics{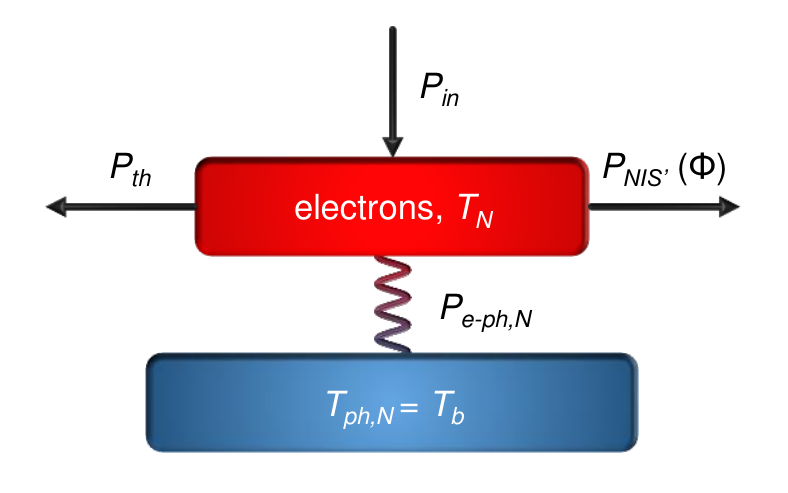}
\caption{\textbf{Thermal model of the normal metal island $N$ with ideal phonon thermalization.} The phonons of the film are fully thermalized with the substrate ($T_{ph,N}=T_b$). The input power injected from the heaters ($P_{in}$) relaxes through the electron-phonon interaction ($P_{e-ph,N}$), the losses through the thermometers ($P_{th}$) and the flux-dependent transport through the $NIS'$ junction [$P_{NIS'}(\Phi)$]. The resulting electronic temperature of the $N$ island ($T_N$) is calculated by the balance equation \ref{BalanceN} and considered homogeneous along all the island. 
} 
\label{FigS1}
\end{figure*}

In the simplest physical picture, the electronic temperature of the normal metal can be assumed to be homogeneous ($T_N\sim$const) and the phonons of $N$ can be considered fully thermalized with the silicon substrate ($T_{ph,N}=T_b$, with $T_{ph,N}$ and $T_b$ the phonon temperature of $N$ and the substrate, respectively), as shown in Fig. \ref{FigS1}. Within these assumptions, the energy balance equation describing the thermal steady state of the $N$ island in the T-SQUIPT reads \cite{Giazotto2006,Fornieri2017}
\begin{align} 
P_{in}=P_{e-ph,N}+P_{th}+P_{NIS'}(\Phi).
\label{BalanceN}
\end{align}
In the above equation, $P_{e-ph,N}=\Sigma_N\mathcal{V}_N(T_N^6-T_{b}^6)$ is the heat current flowing between electrons and phonons in $N$ \cite{Giazotto2006,Taskinen2006}, with $\Sigma_N=4.5\times 10^9$ WK$^{-6}$m$^{-3}$ the electron-phonon coupling constant of Al$_{0.98}$Mn$_{0.02}$ and $\mathcal{V}_N\sim1.7\times10^{-20}$ m$^{3}$ the volume of $N$.
The other contributions to Eq. \ref{BalanceN} ($P_{th}$ and $P_{NIS'}$) are given by the relation for a $NIS$ tunnel junction \cite{Giazotto2006}  
\begin{equation}
P_{NIS}=\frac{1}{e^{2} R_{T}} \int_{-\infty}^{+\infty}  \mathrm{d}E\; E\;\mathcal{N}_S  \left[f_{0}(E,T_{N})-f_{0}(E,T_S)\right] ,
\label{NISpower}
\end{equation} 
where $e$ is the electron charge, $R_T$ is the normal-state tunnel resistance, $T_N$ is the temperature of the normal metal, $T_S$ is the temperature of the superconductor, and
$f_{0}(E,T_i)=\left[1+\exp{\left(E/k_{B}T_i)\right)}\right]^{-1}$
is the Fermi-Dirac distribution at temperature $T_i$ (with $i=N,S$).
The normalized density of states of the superconductor takes the form 
$\mathcal{N}_S (E,T_S)=|\text{Re}\big[(E+i\Gamma)/{\sqrt{(E+i\Gamma)^2-E_g^2(T_S)}}\big]|$, with $\Gamma$ the Dynes parameter accounting for broadening \cite{Dynes1984} and $E_g(T_S)$ the superconducting pairing potential. In particular, $P_{th}$ is the total thermal current flowing between $N$ and the thermometers. It can be calculated by solving Eq. \ref{NISpower} by considering $T_S=T_b$ and the voltage drop $V$ present only for the energized junctions. Finally, the flux-dependent tunnel current flowing from $N$ to the proximitized nanowire [$P_{NIS'}(\Phi)$] can be computed from Eq. \ref{NISpower}, where the density of states of $S'$ ($\mathcal{N}_{S'}$) takes a different form depending on the sweep direction of the magnetic flux \cite{Virtanen2016}. As a consequence, $P_{NIS'}$ is expected to be hysteretic with the direction of $\Phi$.

We now verify the validity of the model by comparing the values of $T_N$ calculated by solving Eq. \ref{BalanceN} for $P_{in}=4.5$ pW with the experimental ones. In particular, the maximum value of $T_N$ is as large as $\sim835$ mK, as shown in the left panel of Fig. 2d of the main text.  In the worst scenario, the gap of $S'$ does not allow any thermal transport across the $NIS'$ junction ($P_{NIS'}=0$) and the thermometers do not dissipate any heat ($P_{th}=0$). Therefore, the input power is only balanced by the energy losses through the phonons ($P_{in}=P_{e-ph,N}$), thus providing the maximum value of $T_N$. Within these assumptions, the model provides a steady-state electronic temperature of $N$ of 625 mK, that is about 200 mK lower than the experimental value of 835 mK. Indeed, the metallic film exchanges less heat that what is expected by this simple thermal model. Therefore, to describe the behavior of the T-SQUIPT, we need to take into account other thermal mechanisms.

\subsection{Finite Kapitza coupling}

\begin{figure*}[th!]
 \centering 
\includegraphics{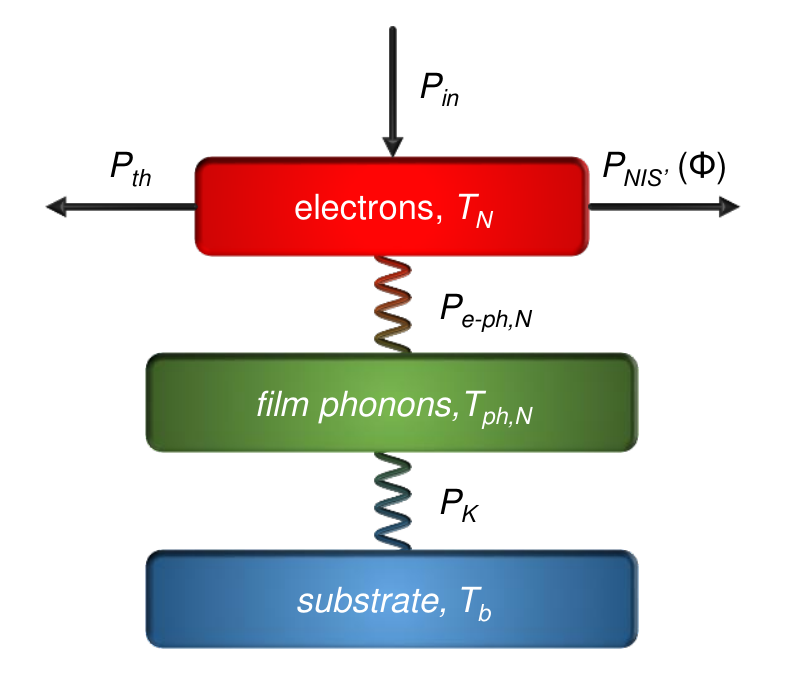}
\caption{\textbf{Thermal model of the normal metal island $N$ with finite Kapitza coupling.} The phonons of $N$ thermalize with the substrate through the Kapitza heat exchange $P_K$, therefore the two temperatures can be different ($T_{ph,N}\neq T_b$). The input power injected from the heaters ($P_{in}$) relaxes through the electron-phonon interaction ($P_{e-ph,N}$), the losses through the thermometers ($P_{th}$) and the flux-dependent transport through the $NIS'$ junction [$P_{NIS'}(\Phi)$]. 
} 
\label{FigS2}
\end{figure*}

Here, we assume a finite thermalization of the phonons of $N$ with the substrates, thus allowing $T_{ph,N}\neq T_b$ (see Fig. \ref{FigS2}). The thermal interaction between the phonons of the normal metal island and the substrate is due to the Kapitza heat exchange
\begin{equation}
P_{K}=\alpha A_N (T_{ph,N}^4 -  T_b^4),
\label{Kapitza}
\end{equation} 
where $\alpha$ is the material dependent coupling constant and $A_N$ is the interface area. 

The thermalization of the $N$ island is limited by electron/phonon or Kapitza coupling depending on the working temperature. In the linear response regime, the electron-phonon and Kapitza thermal conductances are given by
\begin{equation}
    \begin{cases}
    G_{e-ph,N}=6\Sigma_N\mathcal{V}_N T_N^5    \\
    G_K=4 \alpha A_N T_{ph,N}^3.
  \end{cases}
\end{equation}
The crossover temperature ($T^*$) between the two regimes can be calculated by equalizing these two thermal conductivities
\begin{equation}
    \frac{G_{e-ph,N}}{G_K}=\frac{6\Sigma_N\mathcal{V}_N T^{*5} }{4 \alpha A_N T^{*3}}=1.
\end{equation}
The resulting crossover temperature is $T^*=\sqrt{2\alpha/(3\Sigma_Nt_N)}$, where $t_N=15$ nm is the thickness of $N$. In particular, the thermalization follows
\begin{equation}
    \begin{cases}
    T>T^*      & \text{Kapitza limited thermalization}\\
    T<T^*  & \text{electron-phonon limited thermalization},
  \end{cases}
\end{equation}
that is for temperatures larger than $T^*$ the electron-phonon thermalization is more efficient than the Kapitza coupling. 

Since an experimental value of the coupling constant for AlMn is not available, we evaluate $T^*$ by employing the average value for copper thin films evaporated on silicon dioxide $\alpha=50$ pW$\mu$m$^{-2}$K$^{-4}$. Indeed, the Kapitza coupling coefficient of the copper/silicon dioxide interface is in the range $40-60$ pW$\mu$m$^{-2}$K$^{-4}$, that is one order of magnitude lower than the theoretical value. The resulting crossover temperature is $T^*\simeq700$ mK, which is lower than the electronic temperature measured in our experiments. Thus, the Kapitza coupling between the film and substrate phonons limits the thermalization in the T-SQUIPT. We note that the aluminum/silicon interface is theoretically expected to show a lower value of $\alpha$ than the copper/silicon interface. Therefore, the calculated value of $T^*$ overestimates the crossover temperature, thus supporting the hypothesis of overheating of the phonons in $N$ ($T_{ph,N}> T_b$).

Within these conditions, the electronic temperature in $N$ can be calculated by solving:
\begin{equation}
    \begin{cases}
    P_{in}=P_{e-ph,N}+P_{th}+P_{NIS'}(\Phi)\\
    P_{e-ph,N}=P_K,
  \end{cases}
  \label{BalanceTot}
\end{equation}
where the temperatures $T_N$ and $T_{ph,N}$ are the two unknown variables for each value of $T_b$ and $P_{in}$.

\subsection{Model of the real device}

\begin{figure*}[t!]
\includegraphics{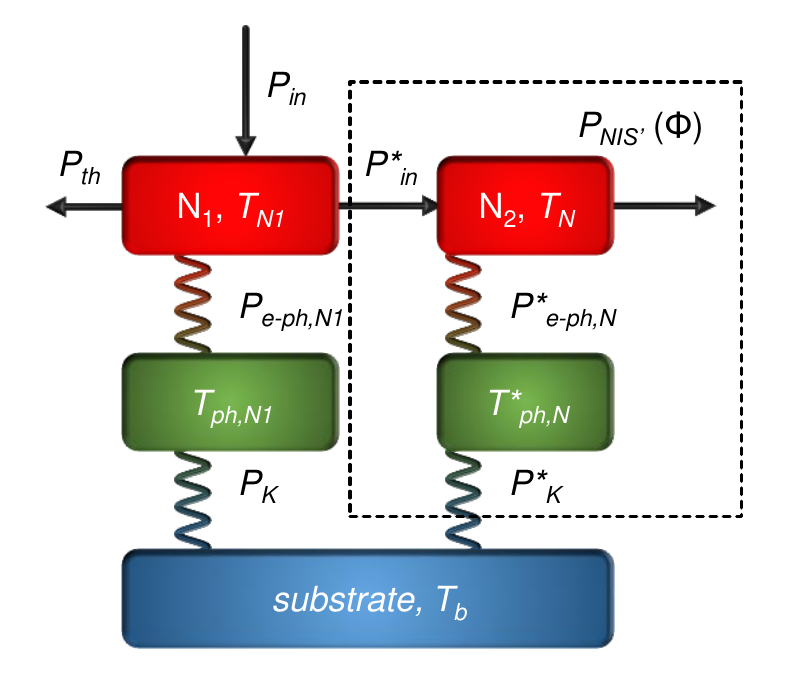}
\caption{\textbf{Thermal model for the real device.} Only a small portion of the $N$ island (indicated with $N_2$, encircled by the black dashed line, is considered in the calculation. This is the portion of the island in contact to the $S'$ weak link. The temperature of this small portion of $N$ can be assumed uniform. We note that this is the area where the temperature is measured by the $SINIS$ thermometer used in the experiment.
The heating power injected from the rest of the island in $N_2$ ($P_{in}^*$) is only a small fraction of the total power injected $P_{in}$ due to the thermal losses to the phonons and the multiple thermometers (see the SEM of the T-SQUIPT in Fig. 1c of the manuscript). 
} 
\label{FigS3}
\end{figure*}

Starting from Eq.~\ref{BalanceTot} it is possible to simulate the behaviour of $T_N(\Phi)$ observed in the experiment. A faithful simulation of the experiment will require a space dependent solution of $T_N$ (the electronic temperature in $N$ is not homogeneous along the long metallic strip). This complex modeling is out of the scope of our analysis. Nevertheless, a good semi-quantitative solution of $T_N(\Phi)$ can be obtained by modeling the heat transport of the small portion of the island included between the thermometer and the $S'$ weak-link where we can assume a uniform temperature. Figure \ref{FigS3} shows a sketch of this simplified thermal model. This portion of the $N$ island ($N_2$) is tunnel coupled to $S'$ towards which it exchanges the flux-dependent heat flow $P_{NIS'}(\Phi)$. Part of the incoming heat will be also dissipated in the phonon bath ($P^*_{e-ph,N}$). The temperature $T_N$ will then rise thanks to the power injected from the rest of the $N$ island ($P_{in}^*$) in good thermal contact with $N_2$. $P_{in}^*$ is lower than the total power injected by the heater ($P_{in}$) and, for simplicity, it  will be re-normalized by a damping factor $d_{loss}$ ($P_{in}^*=P_{in}/d_{loss}$) representing all the thermal losses including the phonon bath and all the $NIS$ junctions (thermometers) in between the heater and thermometer of the device. Therefore, the thermal balance equation for the restricted island can be simplified to:
\begin{equation}
    \begin{cases}
    P^*_{in}=P^*_{e-ph,N}+P_{NIS'}(\Phi)\\
    P^*_{e-ph,N}=P^*_K.
  \end{cases}
  \label{BalanceTot*}
\end{equation}

\begin{figure*}[t!]
\centering 
\includegraphics{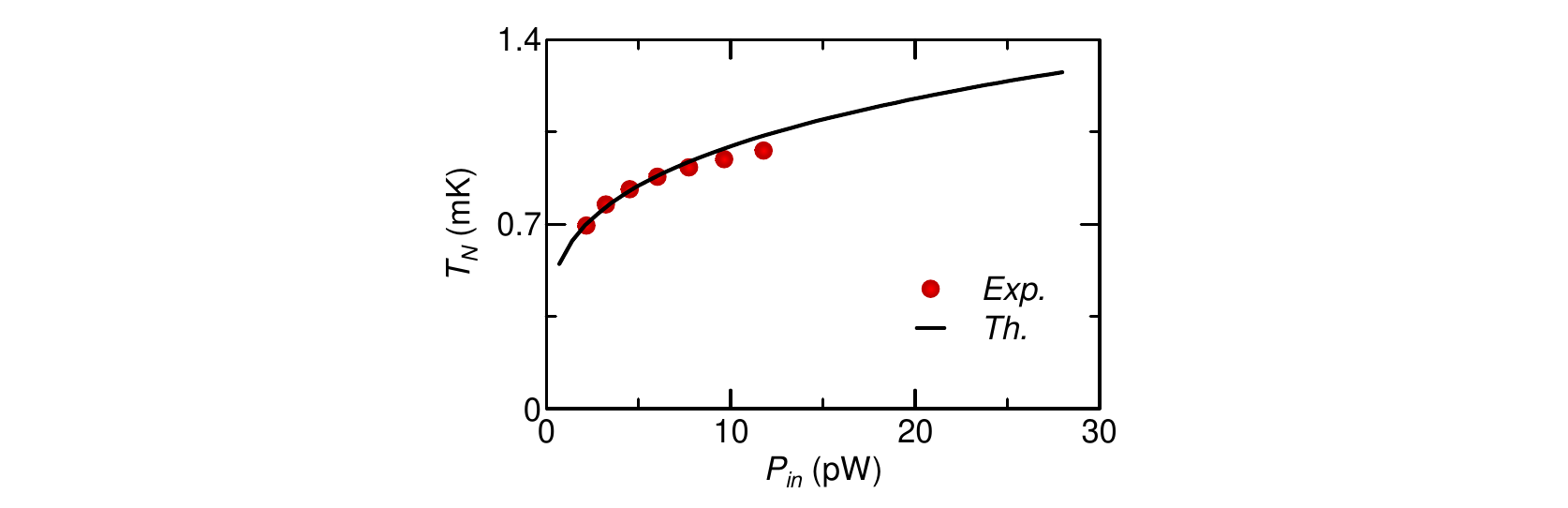}
\caption{\textbf{Fits of $T_N$ vs $P_{in}$.} Comparison between the temperature of the metallic island $T_N$ measured (red dots) and simulated (black line) at zero magnetic flux for different values of heating power $P_{in}$. The same plot is shown in the inset of Fig. 2e in the main text.  
} 
\label{TvsP}
\end{figure*}

From the SEM images of the device (see Fig. 1 of the main text) it is possible to estimate the geometrical dimensions of the island important for the simulation of $P*_{e-ph,N}$ and $P^*_K$. 
A total area of contact with the substrate $A_N=1.125\; \mu$m$^2$ and a volume $\mathcal{V}_N=0.017\; \mu$m$^3$ of $N$ are estimated (with length $l=9\;\mu$m, width $w=125$ nm and thickness $t=15$ nm).
The restricted portion under consideration in the simulation is only a part of this volume. Taking into account a distance of $\simeq 2\;\mu$m between the weak link and the center of the $SINIS$ thermometer, the restricted area of contact and the relative volume are $A^*_N=0.25\; \mu$m$^2$ and $\mathcal{V}^*_N=0.00375\; \mu$m$^3$, respectively.
We simulate $T_N(P_{in})$ by solving Eq. \ref{BalanceTot*} with these parameters. Figure \ref{TvsP} compares the experimental data (red dots) with the simulation evaluated at $\Phi=0$ (continuous black line) and obtained with the following fitting parameters: $\Delta_{0}=200\; \mu$eV, $\Gamma=10^{-3}\Delta_0$, $R_T = 65$ k$\Omega$ extracted from the electrical characterization of the junction while $d_{loss} = 7$, $\alpha=6.25$ pW$\mu$m$^{-2}$K$^{-4}$ have been partially tuned to obtain the best agreement with the experimental data. Notably, the small value of $\alpha$ necessary in the simulation suggests a weak Kapitza coupling for the AlMg respect other $N$ metals like Cu or, alternatively, a less effective area of contact between the substrate and the metallic island.   

\begin{figure*}[t!]
\centering 
\includegraphics{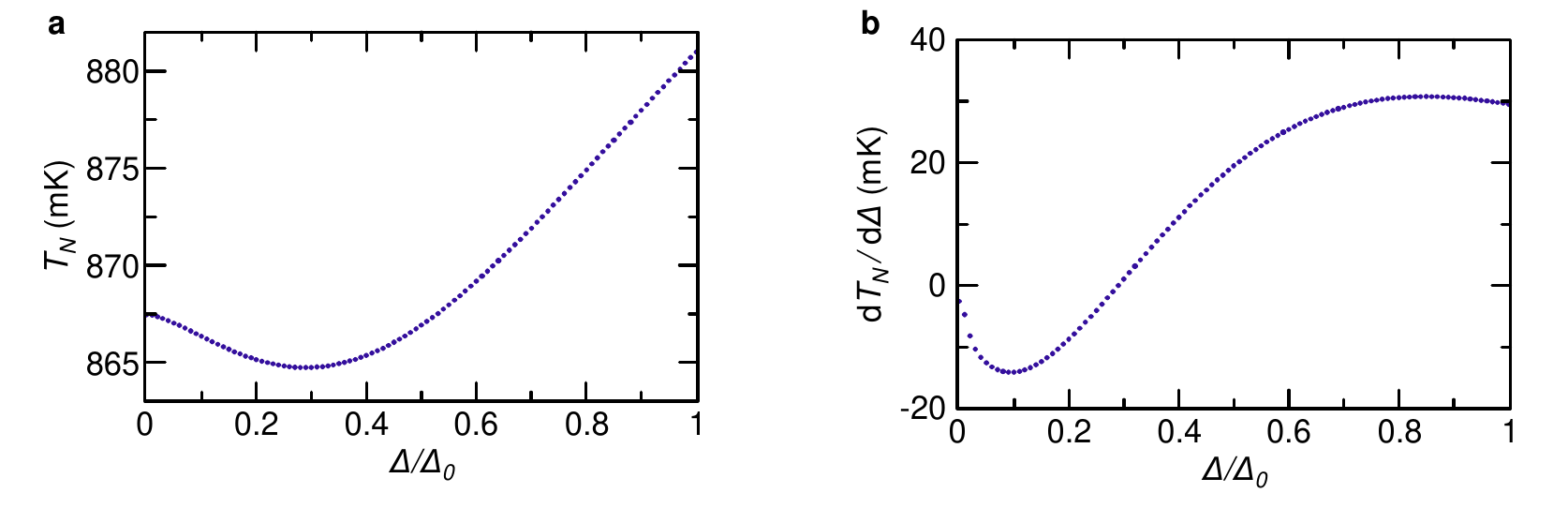}
\caption{\textbf{Evolution of $T_N$ vs $\Delta$.} $T_N$ vs $\Delta$ and its derivative estimated from the Eq.~\ref{BalanceTot*} at base temperature (25 mK) for an injected power of 6 pW. 
} 
\label{FigS4}
\end{figure*}

\begin{figure*}[t!]
\includegraphics{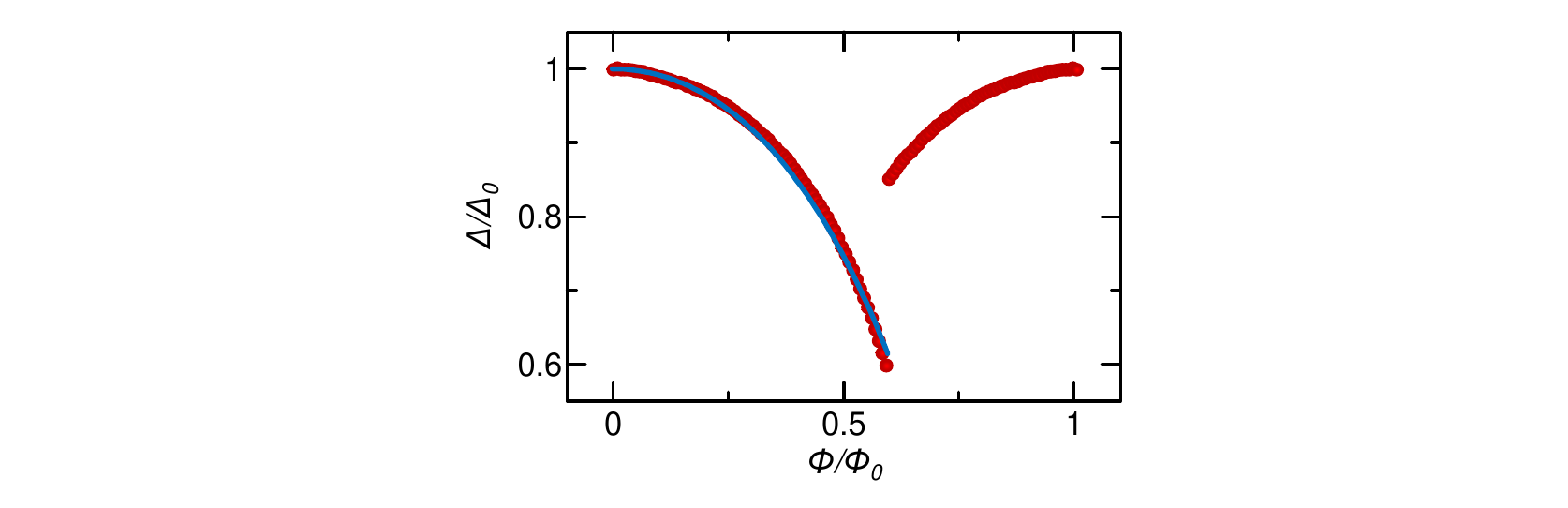}
\caption{\textbf{Superconducting energy gap vs flux.} 
$\Delta/\Delta_0 (\Phi)$ estimated from the mapping of $T_N(\Phi)$ measured at $P_{in}=6$ pW and at $T_b=25$ mK with the linear functional form: $\Delta/\Delta_0 (\Phi) = 1 - (T_N(\Phi)-T_N(0))/0.03\text{ K}$. Data above the phase slip transition at 0.6 $\Phi_0$ have been obtained by mirroring $\Delta(\Phi)$ around $\Phi_0/2$. A bi-quadratic fit of the data (continuous blue line) has been obtained using the approximate formula for $S'$ weak-links \cite{Anthore2003}, $\Delta/\Delta_0 (\Phi)=1-0.75 (\zeta \Phi/\Phi_0)^2 -0.54 (\zeta \Phi/\Phi_0)^4$, with $\zeta= 1.06 $.
} 
\label{FigS5}
\end{figure*}

This fit allows to set most of the model parameters, but to complete the modeling of $T_N(\Phi)$ an accurate estimation of $P_{NIS'}(\Phi)$ is required. This needs a detailed knowledge of the behaviour of the superconducting gap $\Delta(\Phi)$ with the magnetic flux. Experimentally, it could be possible to extract such dependence from the evolution of the gap observed in the tunneling spectroscopy measurements of the junction shown in Fig. 1d of the main text. Still, due to the non-BCS form of the density of states of a long $S'$ weak link \cite{Virtanen2016} this method works without a precise simulation of such density of sates. On the other hand, with the thermal model used in the stimulation we discovered an intriguing linear dependence between $T_N$ and $\Delta$ for the geometry and range of temperatures explored in our experiment. As an example, Fig. \ref{FigS4} shows $T_N$ versus $\Delta/\Delta_0$ estimated from Eq. \ref{BalanceTot*} at base temperature and $P_{in}=6$ pW (panel a) and its derivative (panel b). 
The linear dependence holds
for $\Delta > 0.6\Delta_0$ with a slope of $\simeq 0.03$ K while non-linearities including an upturn at $0.3\Delta_0$ are visible for lower values of $\Delta$. This behaviour is also confirmed for different choices of parameters that modify the slope and the position of the upturn with a non-trivial functional dependence. Thanks to this linear dependence the thermal measurement [$T_N(\Phi)$] can be considered as an accurate mapping of $\Delta(\Phi)$ for modulations of the energy gap above $0.6\Delta_0$. $\Delta(\Phi)$ is then extracted from the experimental $T_N(\Phi)$ measured with an injected power of 6 pW (yellow curve of Fig. 2d of the main text), an is shown in Fig. \ref{FigS5} (this was an arbitrary choice, the other curves show a similar result). Notably, the restricted modulation of the gap [$\Delta(\Phi)<0.6\Delta_0$] guarantees that the linear approximation used in the mapping is correct. Moreover, the evolution of $\Delta(\Phi)$ is also consistent with the evolution of the superconducting gap measured in the tunneling spectroscopy, as shown in Fig. 1d of the main text. An additional sanity check can also be performed by comparing the estimated $\Delta(\Phi)$ with its analytical formula in $\Phi$. The latter can be extracted starting from the analytical approximation of $\Delta$ vs the supercurrent $I_s$ ~\cite{Anthore2003}: 
\begin{equation}
\begin{split}
    \frac{\Delta(\Phi)}{\Delta_0} \simeq 1- 0.75 \Bigl(\frac{\Gamma}{\Delta_0}\Bigr) - 0.54 \Bigl(\frac{\Gamma}{\Delta_0}\Bigr)^2 \\
    \frac{\Gamma}{\Delta_0} \simeq \Bigl(\frac{\sqrt{2} e R \xi_0 I_s}{\pi \Delta_0 L}\Bigr)^2
\label{Anthore}
 \end{split}
\end{equation}
with $R$ the normal-state resistance of the $S'$ weak-link. Considering the almost linear dependence of the current-to-phase relation of long $S'$ junctions $I_s(\Phi)\simeq 10 
\Delta_0 \Phi/e R\Phi_0$~\cite{Virtanen2016}, Eq. \ref{Anthore} can be simplified to: 
\begin{equation}
        \frac{\Delta(\Phi)}{\Delta_0} \simeq 1 - 0.75 \Bigl(\zeta \frac{\Phi}{\Phi_0}\Bigr)^2 - 0.54 \Bigl(\zeta \frac{\Phi}{\Phi_0}\Bigr)^4,
\label{Fit}
\end{equation}

with $\zeta \simeq \frac{10 \sqrt{2}}{\pi} \frac{\xi_0}{L}$. The best fit to the estimated $\Delta(\Phi)$ is obtained for $\zeta\simeq 1$ (see Fig.~\ref{FigS5}) which corresponds to a junction length $L \simeq 4.5 \xi_0$ in good agreement with the length estimated from the electrical characteristics \cite{Ligato2020}.  

All the experimental curves have been fitted with the previous estimation of $\Delta (\Phi),$ as shown in Fig. 2d-e and Fig.3a of the main text. The good agreement between experiment and simulation observed at low temperature (Fig. 2d) confirms the goodness of our model. Furthermore, the approximations made are still valid at intermediate temperatures (see for example the comparison of $T_N(\Phi)$ up to 600 mK in Fig. 3a). At higher temperatures the hysteresis of $\Delta(\Phi)$ have been reduced in accordance to the experimental data and $d_loss$ have been slightly increased ($d_loss(400,600,800 \text{ mK})\simeq 8,9,11$) to adapt the model to the higher $T_b$  and is consistent with and increased coupling between the island and the substrate.  The model is less accurate at higher temperatures, and shows a  reduced temperature modulation  owing to the limitations intrinsic to the chosen approximations.

\end{document}